\documentclass[twocolumn,english,aps,prb,superscriptaddress]{revtex4-2}

\usepackage[T1]{fontenc}
\usepackage[latin9]{inputenc}
\usepackage{amsmath}
\usepackage{graphicx}
\usepackage{xcolor}
\usepackage[colorlinks=true, urlcolor=blue, linkcolor=black, citecolor=black]{hyperref}

\def\eV{{\,\textrm{eV}}}
\def\meV{{\,\textrm{meV}}}
\def\K{{\,\textrm{K}}}

\makeatletter
\usepackage{babel}
\makeatother

\begin{document}

\title{Dzyaloshinskii-Moriya Interaction-Induced Magnetoelectric Coupling
in a tetrahedral Molecular Spin-Frustrated System}
\author{Jie-Xiang Yu}
\affiliation{Department of Physics, Center for Molecular Magnetic Quantum Materials
and Quantum Theory Project, University of Florida, Gainesville, Florida
32611, USA}
\author{Jia Chen}
\affiliation{Department of Physics, Center for Molecular Magnetic Quantum Materials
and Quantum Theory Project, University of Florida, Gainesville, Florida
32611, USA}
\author{Neil Sullivan}
\affiliation{Department of Physics and Center for Molecular Magnetic Quantum Materials, University of Florida, Gainesville, Florida
32611, USA}
\author{Hai-Ping Cheng}
\email{hping@ufl.edu}
\affiliation{Department of Physics, Center for Molecular Magnetic Quantum Materials
and Quantum Theory Project, University of Florida, Gainesville, Florida
32611, USA}

\begin{abstract}

We have investigated magnetoelectric coupling in 
the single-molecule magnet
$\mathrm{Mn}_{4}\mathrm{Te}_{4}(\mathrm{P}\mathrm{Et}_{3})_{4}$ 
with tetrahedral spin frustration.
Our density functional studies found that an electric dipole moment can
emerge with various non-collinear spin orderings.
The forms of spin-dependent dipole are determined and consistent with that in non-centrosymmetric magnets driven by the Dzyaloshinskii-Moriya interaction.
Writing a parameterized spin Hamiltonian, 
after solving for eigenvalues and eigenstates 
we quantified the magnetoelectric coupling by calculating
the thermal average of the electric and magnetic susceptibilities, 
which can be influenced
by external magnetic and electric fields, respectively.
The quadratic relations are expected to be observable in experiments.

\end{abstract}
\maketitle

\section{Introduction}

Considerable interest has been focused recently in the literature  \cite{Cheong_2007a,Spaldin_2010a,Fusil_2014a,Fiebig_2016a,Riviera,Schmid} on the search for multi-functional materials that couple magnetic and electric states through magneto-electric (ME) interactions.
Interest is not only for the fundamental science but also for the possible generation of new electric-field-driven devices and their inherent low power dissipation compared to magnetically driven state changes found in conventional memory and related storage systems. 
In the search for new magnetoelectric materials it is important to note that
ME effects accompany both time-reversal and spatial inversion symmetry
breaking. 
For example, the lattice-mediated ME effect usually happens
when controllable ferroelectric properties without centrosymmetry
coexist with a structure sensitive spin state or spin ordering when
time-reversal symmetry is broken.
The distortion of the lattice influences both electric polarization and magnetic magnetism. 
Besides conventional crystalline
solids, such ME effects based on ionic displacement have also been confirmed in molecule-based magnetic
materials with lower Young's modulus  \cite{Chikara_2019a,Yu_2020a,Zapf_2011a,Yazback_2021a, Chikara, Jakobsen}.

Another origin of ME effect is the non-collinear magnetism characterized
by the Dzyaloshinskii-Moriya (DM) interaction \cite{Dzyaloshinsky1958,Moriya1960}
in non-centrosymmetric magnets. 
In this theory, the polarization is
described by $\hat{\mathbf{e}}_{ij}\times\left(\mathbf{S}_{i}\times \mathbf{S}_{j}\right)$
where $\mathbf{j}_{ij}=\mathbf{S}_{i}\times\mathbf{S}_{j}$ is the so-called spin supercurrent for two spins $\mathbf{S}_{i}$ and $\mathbf{S}_{j}$
and $\hat{\mathbf{e}}_{ij}$ is the unit vector connecting the two
spins \cite{Katsura_2005a,Katsura_2007a,Cheong_2007a}. 
The mechanism
of DM-induced ME effect is confirmed in some spiral magnetic system
such as rare-earth manganite $\mathrm{TbMnO}_{3}$ and $\mathrm{DyMnO}_{3}$  \cite{Kimura_2007a}.
Although experiments found non-structural induced ME effects in some
polynuclear molecular nanomagnets \cite{Boudalis_2018a,Robert_2019a},
DM-induced ME effect studies in molecular magnets remain largely
under-investigated. 
Special quantum features of the quantum spin states
in molecular magnets that differentiate these systems from other crystalline
materials can provide new ME couplings with potential applications
in quantum information science.

\begin{figure}
\includegraphics[width=1\columnwidth]{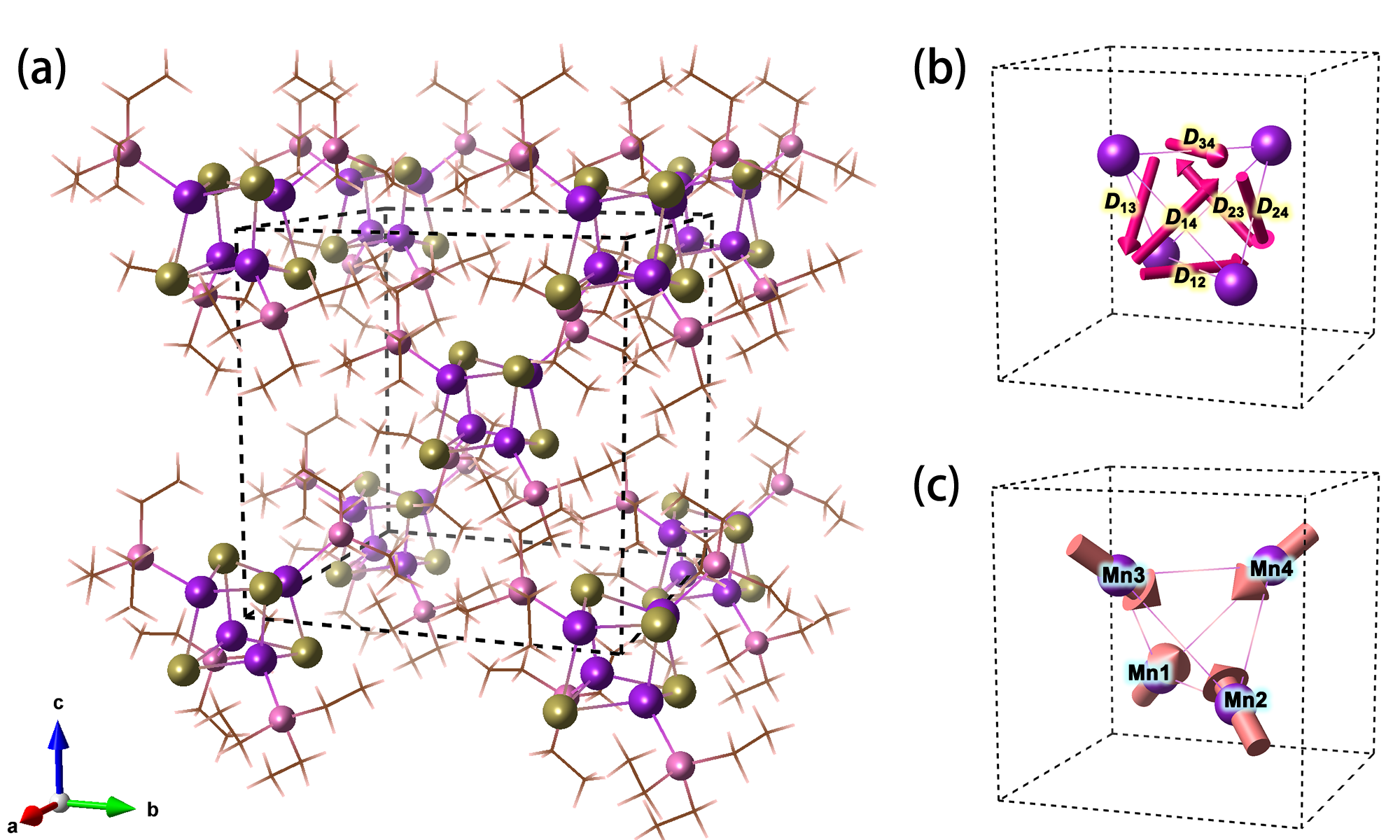}

\caption{(a) The crystalline phase of $\mathrm{Mn}_{4}\mathrm{Te}_{4}(\mathrm{P}\mathrm{Et}_{3})_{4}$.
Purple: Mn, dark yellow: Te, pink: P. 
(b) The DM vectors and (c) the axis of magnetic anisotropy in $\mathrm{Mn}_{4}\mathrm{Te}_{4}(\mathrm{P}\mathrm{Et}_{3})_{4}$.}
\label{fig:Mn4-dm-Ku} 
\end{figure}

When we consider the symmetry of a molecular magnet, tetrahedral symmetry
with point group $T$ (chiral tetrahedral symmetry) is a rare example
in which the absence of spatial inversion symmetry alone 
does not bring about a net polarization. 
Furthermore,
an antiferromagnetic exchange interaction in a tetrahedral geometry can lead to frustrated
spins, where the ground spin state can be uncertained and easily be altered by external 
fields. A representative multiferric crystal system
with tetrahedral structure is $\mathrm{Cu}_{2}\mathrm{O}\mathrm{SeO}_{3}$
which hosts magnetically induced polarization in the ferrimagnetic,
helimagnetic, and skyrmion crystal phases  \cite{Yang_2012a,Seki_2012a,Seki_2012b,White_2014a,Ruff_2015a}
because of DM interaction.
In contrast to the distorted $\mathrm{Cu}_{4}$
tetrahedron that does not respect tetrahedral spin frustration, the
$\mathrm{Mn}_{4}\mathrm{Te}_{4}\left(\mathrm{P}\mathrm{Et}_{3}\right)_{4}$ 
molecule 
where the magnetic center $\mathrm{Mn}_{4}$ forms an equilateral
tetrahedron in this study is a magnetically frustrated unit.
The crystalline phase of $\mathrm{Mn}_{4}\mathrm{Te}_{4}(\mathrm{P}\mathrm{Et}_{3})_{4}$
shown in Fig.~\ref{fig:Mn4-dm-Ku} has a body-centered-cubic lattice
and the space group is $I23(197)$ with point group $T$, so that
both the global symmetry and local chemical environment respect a
perfect tetrahedral symmetry.

In this article, we have investigated both the magnetic
properties and the electric polarization for various spin states of $\mathrm{Mn}_{4}\mathrm{Te}_{4}(\mathrm{P}\mathrm{Et}_{3})_{4}$ based
on density functional calculations. We confirmed the DM-induced ME
effect in the molecular magnets. After solving the eigenvalues and
eigenstates of the parameterized spin Hamiltonian, we quantified the
magnetoelectric coupling by calculating the thermal average of the
electric susceptibility, which can be influenced by external magnetic
field. The rest of the paper is organized as follows: In Section II,
we describe the computational details; in Section III, we present
results from DFT calculation and model Hamiltonian; 
and finally in Section IV, we conclude our investigation.

\section{Computational methods}

Our density functional theory (DFT)-based calculations are performed
with projector augmented wave pseudopotentials  \cite{PAW_1994,PAW_1999}
implemented in the Vienna ab initio simulation package 
(VASP) \cite{Kresse_1996_CMS,Kresse_1996_PRB}.
The generalized gradient approximation (GGA) in the Perdew,
Burke, and Ernzerhof (PBE) formation \cite{PBE} is used as the exchange-correlation energy, 
and the Hubbard $U$ method $(U=4.0\eV$, $J=0.9\eV$) with
density only and a spin-independent double counting scheme \cite{Anisimov_1993a} 
is applied on Mn($3d$) orbitals to include strong-correlation effects.
An energy cutoff of $600\eV$ is used for the plane-wave expansion throughout
the calculations. The DFT-D3 method \cite{DFT_D3_2010a} with inclusion
of van der Waals correction is employed. For non-collinear spin orderings, 
spin-orbit couplings (SOC) are included. The polarization vectors were
obtained by the evaluation of the Berry phase expressions  \cite{Polar_KingSmith_1993,Polar_Resta_1994a}.

We use a body-centered cubic lattice with experimental lattice constant
13.174\AA \cite{Choi_2015a} including one $\mathrm{Mn}_{4}\mathrm{Te}_{4}(\mathrm{P}\mathrm{Et}_{3})_{4}$
molecule for all calculations. The $K$-points were sampled on a $7\times7\times7$ 
$\Gamma$-centered mesh in the Brillouin zone.

\section{Results}

\subsection{Density functional results}

\begin{figure}
\includegraphics[width=1\columnwidth]{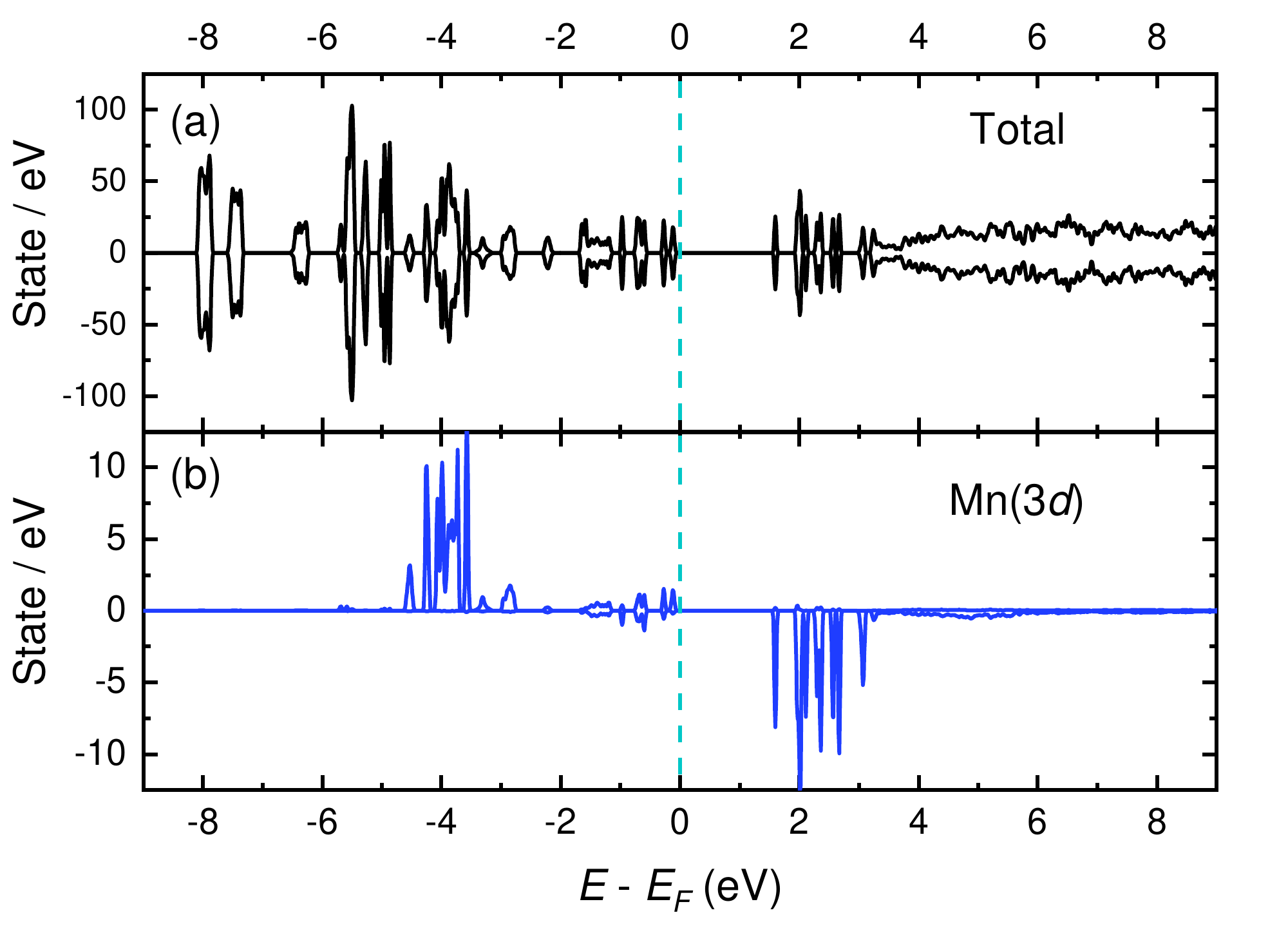}

\caption{Without spin-orbit coupling: (a) the total density-of-states of $\mathrm{Mn}_{4}\mathrm{Te}_{4}(\mathrm{P}\mathrm{Et}_{3})_{4}$
in the two-up-two-down spin configuration. 
(b) The PDOS for Mn($3d$) orbitals. Positive and negative values refer to spin-majority channel and spin-minority channel,  respectively. 
The Fermi energy is set to zero.}
\label{fig:electron} 
\end{figure}

DFT results showed that the local magnetic spin moment on each Mn
is $4.27\mu_{B}$ in a collinear spin configuration where two of
four Mn spins are up and other two are down, without spin-orbit coupling.
As shown in Fig.~\ref{fig:electron}(a), the total density-of-states
(DOS) has a gap about $1.5\eV$, indicating an insulating nature. The
corresponding projected density-of-states (PDOS) results (see Fig.~\ref{fig:electron}(b))
show that all Mn($3d$) components in the spin-majority channel are fully
occupied while almost all Mn($3d$) components in the spin-majority are above the Fermi energy. Thus, each Mn ion has five spin-up electrons
half filling the $d$ orbitals, following Hund's rule, and 
in $\mathrm{Mn}_{4}\mathrm{Te}_{4}(\mathrm{P}\mathrm{Et}_{3})_{4}$,
each $\mathrm{Mn}$ displays a $+2$ valence state and $S=5/2$ high spin
state. Before building a spin Hamiltonian, we point out that our calculations indicate that in this system the strain effect is negligible.

Because of the absence of inversion symmetry, the exchange interaction
between two local magnetic spins on Mn includes an off-diagonal contribution, the Dzyaloshinskii-Moriya (DM) interaction \cite{Dzyaloshinsky1958,Moriya1960}.
The strength of the DM interaction is proportional to the strength of
SOC of the bridging Te, and is not negligible. The spin-spin
Hamiltonian that properly includes this interaction for the four $S=5/2$ 
spins on $\mathrm{Mn}^{2+}$ ions with tetrahedral symmetry is given
by:
\begin{equation}
\mathcal{H}_{0} =\sum_{\left\langle i,j\right\rangle } \left[J \, \mathbf{S}_{i}\cdot\mathbf{S}_{j}+\mathbf{D}_{ij}\cdot\left(\mathbf{S}_{i}\times\mathbf{S}_{j}\right)\right]
  -K_{u}\sum_{i}\left(\mathbf{M}_{i}\cdot\mathbf{S}_{i}\right)^{2}
\label{eq:h0}
\end{equation}
where $J$ is the Heisenberg interaction and $\mathbf{D}_{ij}=D \, \hat{\mathbf{D}}_{ij}$
is the DM vector for two local spins on neighboring $\mathrm{Mn}$ sites
$i$ and $j$. Following Moriya's rule \cite{Moriya1960}, 
the direction $\hat{\mathbf{D}}_{ij}$ is perpendicular to the Mn-Mn bond (see Fig.~\ref{fig:electron}(b)). $K_{u}$ is the magnitude of magnetic anisotropy
(note that $K_{u}$ is often denoted as $D$ in molecular magnet literature)
and $\mathbf{M}_{i}$ is a unit vector which represents the direction
of the magnetic anisotropy on $\mathrm{Mn}$ site $i$. Because of
the tetrahedral symmetry, $\mathbf{M}_{i}$ is directed from $\mathrm{Mn}$ site $i$ to the body center of tetrahedron
(see Fig.~\ref{fig:electron}(c)).

\begin{figure}
\includegraphics[width=1\columnwidth]{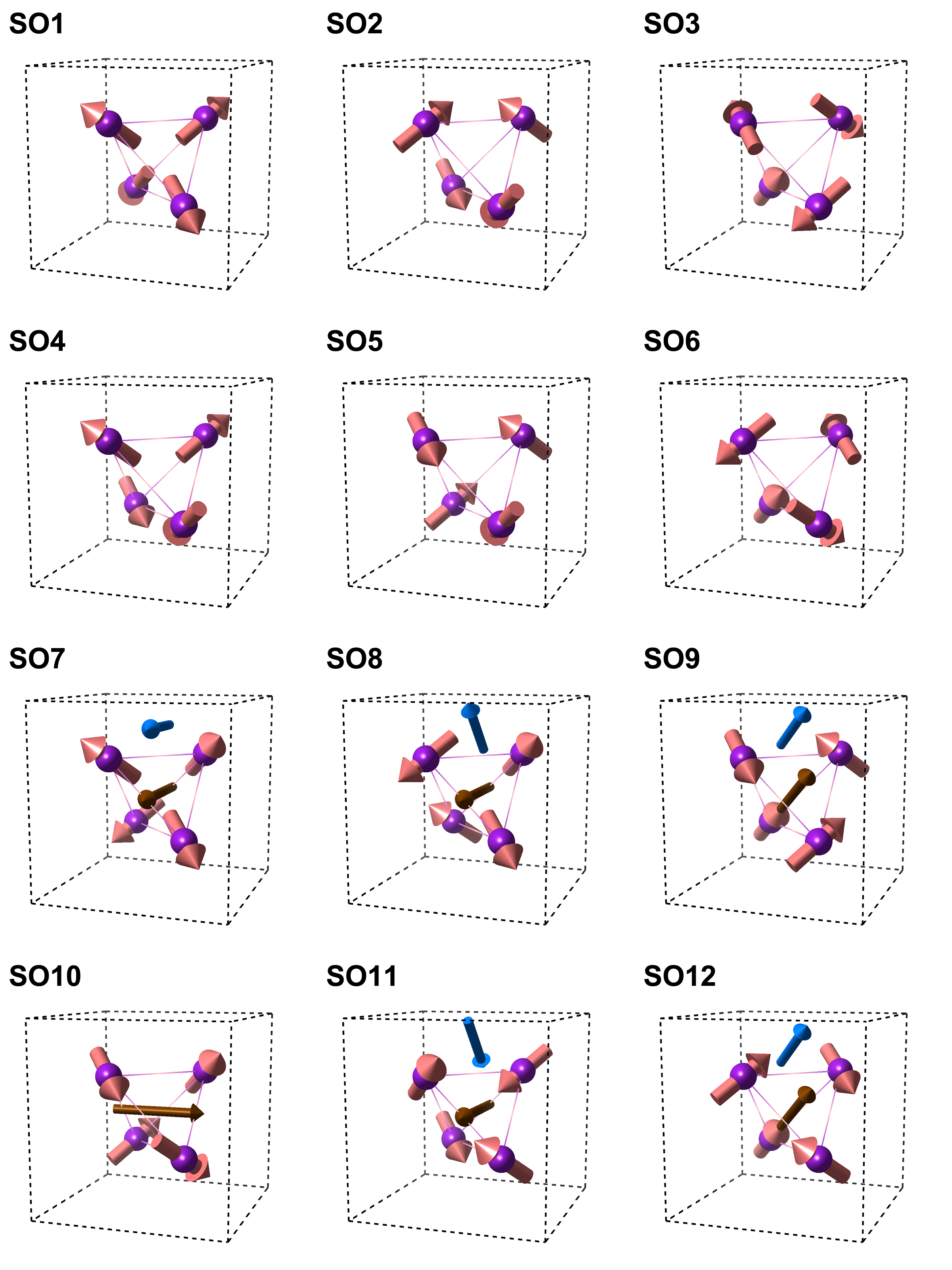}
\caption{Twelve non-collinear spin configurations, labeled SO1 to SO12,
used for total energy calculations and electric dipole calculations.
Brown and blue arrows corresponds to the directions of the total magnetization
and the electric dipole moment for each spin configuration, respectively. }
\label{fig:SO-all} 
\end{figure}

Based on the Hamiltonian, we investigate the magnetic properties of
$\mathrm{Mn}_{4}\mathrm{Te}_{4}(\mathrm{P}\mathrm{Et}_{3})_{4}$
by calculating total energies for two collinear spin configurations
along the $[111]$ direction of the cubic crystalline lattice,
all-up and two-up-two-down, and twelve non-collinear spin configurations, labeled SO1 to SO12, including six zero magnetization configurations
and six non-zero magnetization configurations, shown in Fig.~\ref{fig:SO-all}.
The relative total energies are listed in Table~\ref{tab:so-all}. 
We transfer spin
configurations into quantum spin states. For each spin configuration
$\alpha$, each local magnetic spin $i$ on $\mathrm{Mn}^{2+}$ has
a normalized classical spin vector $\mathbf{e}_{i}=(e_{ix},e_{iy},e_{iz})$
and the spin quantum number $s_{i}=5/2$. Diagonalizing the spin matrix
$\mathbf{e}_{i}\cdot\mathbf{S}_{i}$ where $\mathbf{S}_{i}$ is the
matrix of the spin operator, we obtain the quantum spin state for
this spin in the basis of $\left|s_{iz}\right\rangle $ which is the
eigenvector $\left|\alpha_{i}\right\rangle $ with the eigenvalue
$+5/2$. Therefore, the quantum spin state for this spin configuration
is $\left|\alpha\right\rangle =\left|\alpha_{1}\right\rangle \otimes\left|\alpha_{2}\right\rangle \otimes\left|\alpha_{3}\right\rangle \otimes\left|\alpha_{4}\right\rangle $.
Then the super-rank linear equations based on all spin configurations
are 
\begin{equation}
 \{E_{\alpha} = E_{0}+\bigl\langle
\alpha\bigm|\mathcal{H}_{0}\bigm|\alpha\bigr\rangle\}_{\alpha}
\end{equation}
\begin{eqnarray}
\bigl\langle \alpha\bigm|\mathcal{H}_{0}\bigm|\alpha\bigr\rangle &=& J\sum_{\bigl\langle i,j\bigr\rangle }\bigl\langle \alpha\bigm|\mathbf{S}_{i}\cdot\mathbf{S}_{j}\bigm|\alpha\bigr\rangle 
\label{eq:fitting} \\
&&\quad +D\sum_{\bigl\langle i,j\bigr\rangle }\bigl\langle \alpha\bigm| \hat{\mathbf{D}}_{ij}\cdot\left(\mathbf{S}_{i}\times\mathbf{S}_{j}\right)\bigm|\alpha\bigr\rangle\nonumber \\
&&\quad -K_{u}\sum_{i}\bigl\langle \alpha\bigm|\left(\mathbf{M}_{i}\cdot\mathbf{S}_{i}\right)^{2}\bigm|\alpha\bigr\rangle 
\nonumber
\end{eqnarray}
where $E_{\alpha}$ is the total energy of spin configuration $\alpha$
and $E_{0}$ is the spin-irrelevant energy. The parameters $J$, $D$, and $K_{u}$ are then obtained from energy fitting.
According to a suggestion of Ruiz \textit{et al.} \cite{Ruiz_1999a,Ruiz_2003a},
when the spin broken symmetric antiferromagnetic spin ordering includes
overlapped occupied molecular orbitals, a correction of $1+{\min\left(S_{i},S_{j}\right)}/{2S_{i}S_{j}}$
is included for obtaining the expectation values of the exchange interaction terms.
The solution according to a least-square fit leads to $J=14.02\meV$, $D=-0.44\meV$, and $K_{u}=0.26\meV$. 
Positive $J$ indicates antiferromagnetic coupling.
A negative $D$ gives an opposite direction of spiral direction compared to a positive one. Positive $K_{u}$ means
that each $\mathbf{M}_{i}$ of site $i$ is an easy axis.

\begin{table}[h]
\caption{The relative total energies and the magnitude of the dipole moment for each spin configuration. 
SO1-SO12 correspond to the non-collinear spin
configurations shown in Fig.~\ref{fig:SO-all}. CO-uudd and CO-uuuu
correspond to two collinear configuration along the $[111]$ 
direction of the cubic crystalline lattice, all-up and two-up-two-down, respectively.}
\begin{ruledtabular}
\begin{tabular}{crr}
 & $E$ (meV) & $P$ ($e$\AA) \\
\hline 
SO1 & 0.00 & 0.000 \\
SO2 & $-20.80$ & 0.000 \\
SO3 & $-17.32$ & 0.000 \\
SO4 & $-24.27$ & 0.000 \\
SO5 & $-9.17$ & 0.000 \\
SO6 & $-17.3$ & 0.000 \\
SO7 & 275.94 & 0.012 \\
SO8 & 265.13 & 0.019 \\
SO9 & 208.24 & 0.058 \\
SO10 & 276.83 & 0.000 \\
SO11 & 265.13 & 0.019 \\
SO12 & 208.24 & 0.058 \\
CO-uudd & 12.11 & 0.000 \\
CO-uuuu & 946.43 & 0.000 \\
\end{tabular}
\end{ruledtabular}
\label{tab:so-all}
\end{table}


The electric dipole moments are calculated for each spin configuration.
According to the results shown in Table.~\ref{tab:so-all}, 
both collinear spin configurations and all six non-collinear configurations with zero magnetization have zero dipole moment.
Among non-collinear
configurations from SO7 to SO12, SO7 has a non-zero electric dipole
moment $0.012 \, e\textrm{\AA}$ along the $[100]$ direction, the same
as the direction of its net magnetization.
Spin configurations SO8 and SO11, with the same magnetization along the $[100]$ direction, provide an electric dipole of 
the same magnitude $ 0.019 \,e \textrm{\AA}$ and 
opposite orientations along the $[101]$ direction.
Both SO9 and SO12 retain a three-fold rotation axis along the $[111]$
direction with non-zero magnetization and dipole moment 
$ 0.058 \, e\textrm{\AA}$ along the same direction.
Configuration SO10 has zero electric dipole moment. 
All dipole moments are plotted in Fig.~\ref{fig:SO-all}.
Note that the atomic positions as well as the lattice are fixed, so
that the calculated dipole moments are purely from charge density displacement 
driven by non-collinear spin ordering, indicating a magnetoelectric
coupling. The dipole moments are not changed significantly when the
atomic positions are relaxed in their spin configurations.
The spin-driven magnetoelectric coupling is robust.

The spin-dependent electric dipole moments do not change magnitude
or sign when all spins are reversed, indicating that the dipole is a function
of even order in spins.
SO11 is the spin configuration where spins
on Mn1 and Mn2, Mn3 and Mn4 are exchanged from SO8, reversing 
the direction of the dipole moment. Furthermore, SO9 and SO12, with
opposite spin chiralities, result the same dipole moment, so the
dipole moment is not relevant to chiral spin textures. Based on the
dipole moment results from DFT and analysis based on symmetry properties, 
we obtain the spin-dependent electric dipole moment as a function
of spins as, 
\begin{equation}
\mathbf{P}=\alpha\sum_{\left\langle i,j\right\rangle }\hat{\boldsymbol{e}}_{ij}\times\left(\mathbf{S}_{i}\times\mathbf{S}_{j}\right)
\label{eq:dipole}
\end{equation}
where $\hat{\mathbf{e}}_{ij}$ is the direction from site $i$ to $j$. 
The magnitude of the coefficient $\alpha$ is about $0.005$ --  $0.035\,e\textrm{\AA}$ in $\mathrm{Mn}_{4}\mathrm{Te}_{4}(\mathrm{P}\mathrm{Et}_{3})_{4}$ based on different spin configurations with non-zero spin-dependent electric dipole moment.
Here we demonstrate that even at the single molecular scale, the DM-induced ME effect is still valid. 

We also investigated the magnetic and dielectric properties under hydraulic external strain by modulating the lattice constant. 
As a result, both no significant response of the relative total energies is identified, 
and dipole moments for the spin configurations are almost invariant. 
This indicates that the ME effect are insensitive to strain. 

\subsection{Quantum spin model}


Once the spin-dependent electric dipole moment is determined, the
Hamiltonian for the response to external magnetic field and electric
field is given by 
\begin{equation}
\mathcal{H}=\mathcal{H}_{0}-\mathbf{B}\cdot\sum_{i}\mathbf{S}_{i}-\mathbf{E}\cdot\mathbf{P}\label{eq:h}
\end{equation}
where $bm{E}$ is the electric field, $\mathbf{P}$ is the spin-driven
polarization, and $\mathbf{B}=g\mu_{B}\mu_{0}\mathbf{H}$ is proportional
to the magnetic field $\mathbf{H}$. The electric field is coupled
with spins since the electric dipole moment is a function of spin 
as in eq.~(\ref{eq:dipole}), with $\alpha=0.035\, e\textrm{\AA}$ chosen.
We diagonalized the Hamiltonian matrix for various $\mathbf{B}$ and $\mathbf{E}$ and obtained the total $6^{4}=1296$ eigenvalues and eigenstates.
The corresponding quantum spin states and the expectation values of polarization are also obtained.

\begin{figure}
\includegraphics[width=1\columnwidth]{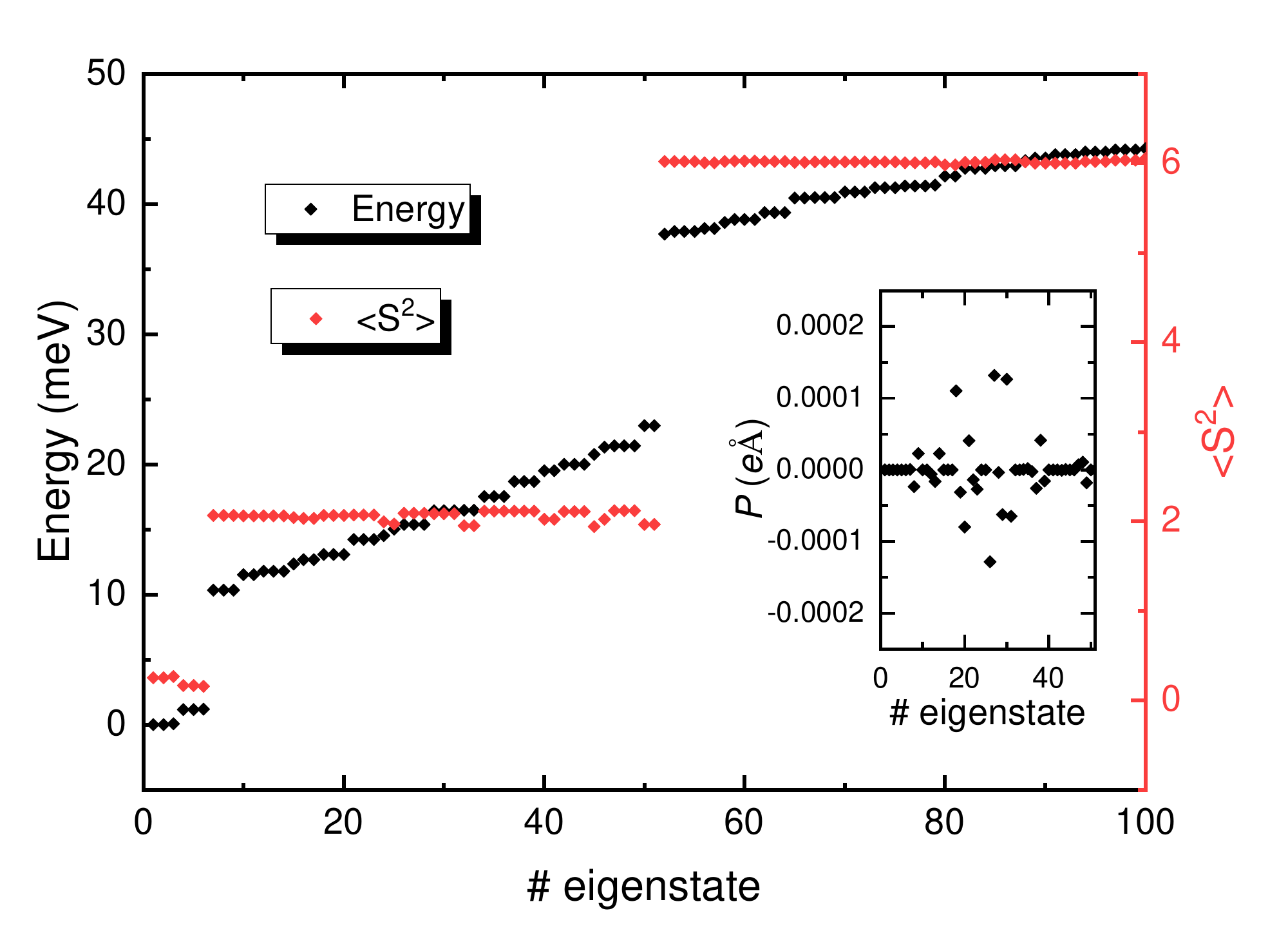}

\caption{Without external magnetic and electric fields, the eigenvalues (energies)
and the expectation value of $\langle S^{2}\rangle $ of
the first 100 eigenstates. The energy of the ground state is set to
zero. Insert: the expectation value of the dipole moment for the first
51 eigenstates. }

\label{fig:Q-spin}
\end{figure}

The eigenvalues and the expectation value of $\langle S^{2}\rangle $ of the first 100 eigenstates Under zero magnetic and electric field 
are shown in Fig.~\ref{fig:Q-spin}.
Note that because of the DM interaction in the Hamiltonian, 
the total spin $S$ of the molecule is not a good quantum number 
and the expectation value of $\langle S^{2}\rangle $ 
is not precisely $S(S+1)$ for each eigenstate.
However, since $D\ll J$ in 
$\mathrm{Mn}_{4}\mathrm{Te}_{4}(\mathrm{P}\mathrm{Et}_{3})_{4}$,
the integer spin quantum number can still be used to label the spin
states. The first six eigenstates with the lowest energies have  $\langle S^{2}\rangle $ close to zero, so that the these states correspond to a $S=0$ quantum spin state.
The next 45 states, which are about $10\meV$ higher than the 
$S=0$ states, have $\langle S^{2}\rangle $ near 2, 
corresponding to a $S=1$ state. The final 49 eigenstates are about
$40\meV$ higher than $S=0$ states and have $\langle S^{2}\rangle $
near 6, corresponding to a $S=2$ state. 
Considering that the energy scale
of external fields is several meV, we focus on $S=0$ and $S=1$
states. 

The results of expectation values of polarization for $S=0$ and $S=1$
states are shown in the insert of Fig.~\ref{fig:Q-spin}. All six of the $S=0$ states have zero dipole moment.
Some $S=1$ states have a non-zero polarization, but the magnitude of the dipole moment is much smaller than the non-zero dipole obtained from DFT calculations.
It is because that according to the DFT results, the spin configurations with non-zero dipole such as SO9 and SO12 have non-zero total magnetization and are more than $200\meV$ higher in energy than the spin configurations with zero magnetization.
Therefore, the quantum spin states $S=1$, the superposition of classical spin configurations, are dominated by zero magnetization configurations and only have very small non-zero dipole moment. 

\begin{figure}
\includegraphics[width=1\columnwidth]{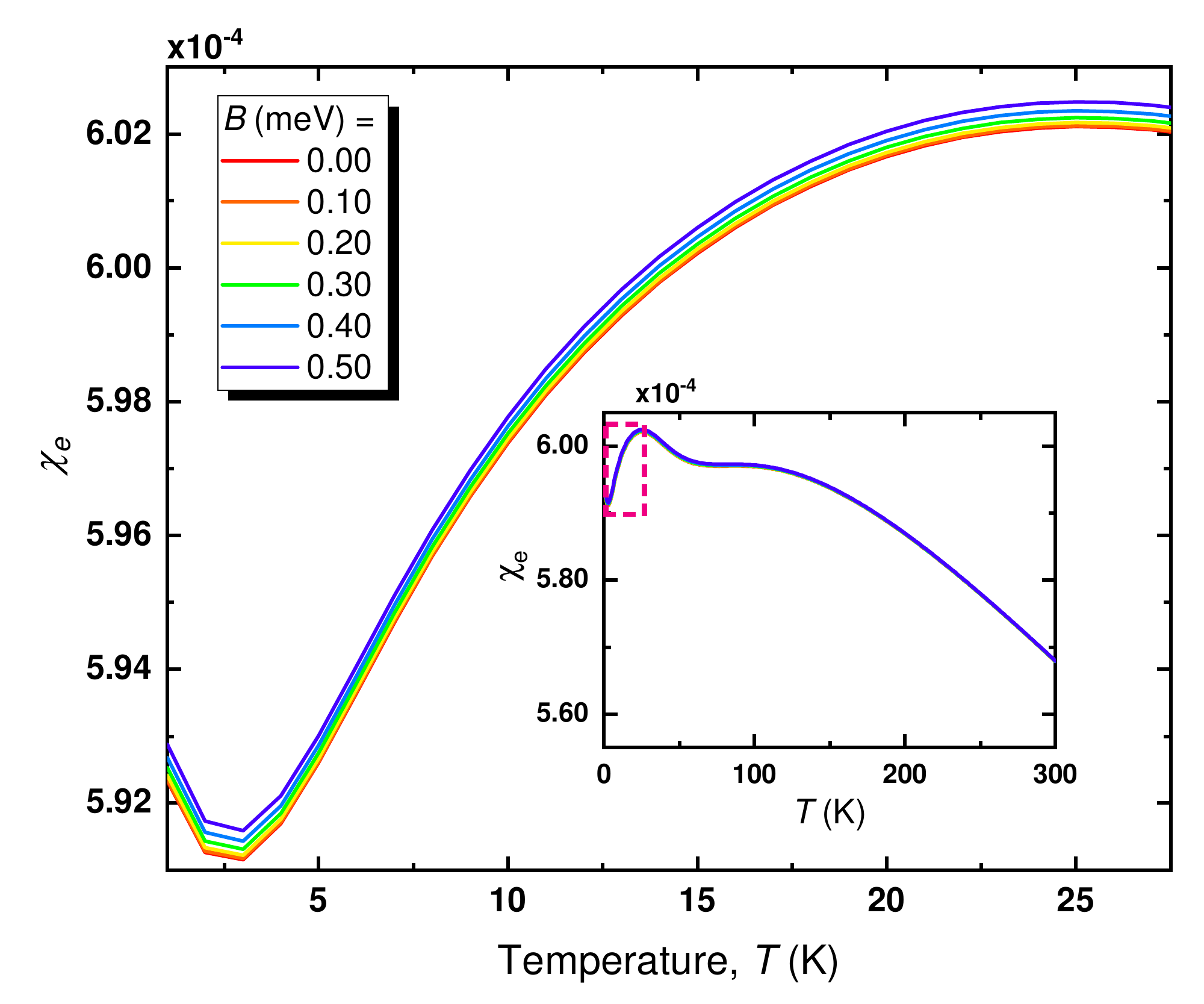} 
\caption{The electric susceptibility  
as a function of temperature with various magnetic fields. 
The red dashed rectangle in the insert panel shows the region expanded in the main panel.
The direction of electric susceptibility is perpendicular to the direction of magnetic field (along $c$-axis in Fig.~\ref{fig:Mn4-dm-Ku}(a)).
}
\label{fig:thermal-suse} 
\end{figure}

\begin{figure}
\includegraphics[width=1\columnwidth]{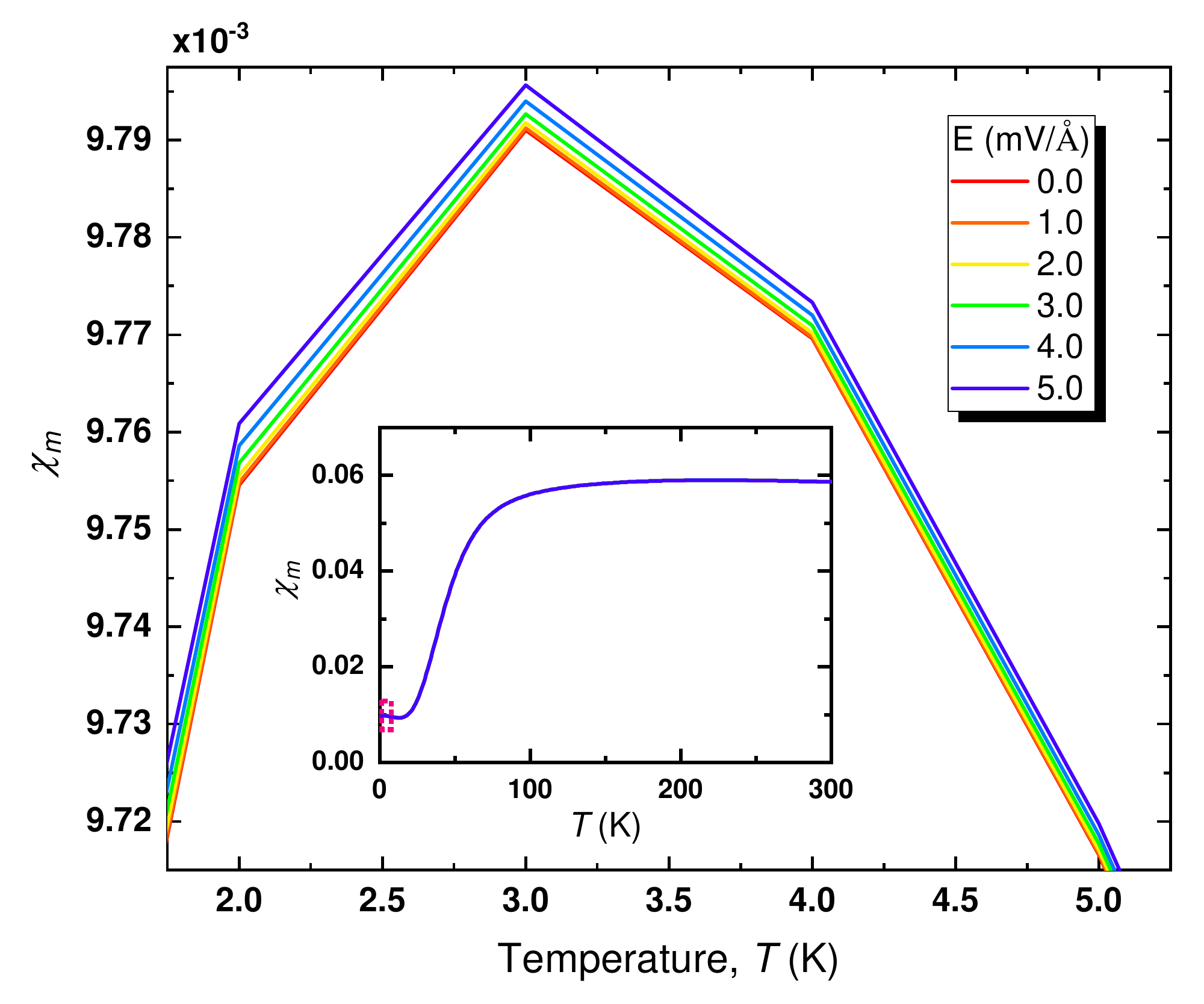} 
\caption{The  magnetic susceptibility as a function
of temperature with various electric fields.
The red dashed rectangle 
in the insert panel shows the region expanded in the main panel.
The direction of magnetic susceptibility is perpendicular to the direction of electric fields (along $b$-axis in Fig.~\ref{fig:Mn4-dm-Ku}(a)).
}
\label{fig:thermal-susm} 
\end{figure}

Based on the eigenvalues, eigenstates, and the corresponding expectation
values of spins and dipoles of quantum spin model, we obtained 
thermal properties of that system. 
The corresponding partition function 
$Z$, thermal average of magnetization $\bar{\mathbf{m}}$ and dipole
$\bar{\mathbf{P}}$ at finite temperature $\beta=1/k_{B}T$ are given
by 
\begin{eqnarray}
Z(\mathbf{E},\mathbf{B},\beta) &=& \sum_{i}\exp\left(-\beta\varepsilon_{i}\right)\\
\bar{\mathbf{m}}\left(\mathbf{E},\mathbf{B},\beta\right) &=& \frac{g\mu_{B}}{Z}\sum_{i}\left\langle \mathbf{S}\right\rangle \exp\left(-\beta\varepsilon_{i}\right)\\
\bar{\mathbf{P}}\left(\mathbf{E},\mathbf{B},\beta\right) &=& \frac{1}{Z}\sum_{i}\left\langle \mathbf{P}\right\rangle \exp\left(-\beta\varepsilon_{i}\right)
\label{eq:thermal}
\end{eqnarray}
where the summation is over all eigenvalues $\{ \varepsilon_{i}\} $.
Then the corresponding electric susceptibility $\chi_{e}$ which depends on magnetic fields is given by 
\begin{equation}
\chi_{e}\left(\mathbf{E},\mathbf{B},\beta\right)  =\frac{\partial\mathbf{\bar{P}}\left(\mathbf{B},\mathbf{E}\right)}{\partial\mathbf{E}}
 . \label{eq:sus_e}
\end{equation}
Similarly, the magnetic susceptibility  $\chi_{m}$ influenced by electric fields is given by 
\begin{equation}
\chi_{m}\left(\mathbf{E},\mathbf{B},\beta\right)  =\frac{\partial\bar{\mathbf{m}}\left(\mathbf{B},\mathbf{E}\right)}{\partial\mathbf{B}}
 . \label{eq:sus_m}
\end{equation}
The temperature dependent results for $\chi_{e}$ and $\chi_{m}$ are shown in Fig.~\ref{fig:thermal-suse} and Fig.~\ref{fig:thermal-susm}, respectively. 
In the inserts for both $\chi_{e}$ and $\chi_{m}$, 
dashed contours identify the region of non-zero magnetoelectric response, where $\chi_{e}$ can be affected by magnetic fields and $\chi_{m}$ is modulated by the electric fields, though the magnitude of magnetoelectric coupling is very small.

All the electric susceptibility curves have a local minimum at $T\sim3$K and a local maximum at $T\sim25$K.
Meanwhile, all the magnetic susceptibility curves have a local maximum about at $T\sim3$K and a local minimum at $T\sim25$K. 
Since the energy of about $2.2\meV$ corresponding to $25\K$ is much smaller than the energy gap between $S=0$ and $S=1$ states, the thermal average of $\chi_{e}$ and $\chi_{m}$ below $25\K$ is  determined by only the first six $S=0$ states. 
For each eigenstate $i$, the contribution to $\chi_{m}$ is proportional to the fluctuation of spins 
$\langle \mathbf{S}^2\rangle - \langle \mathbf{S}\rangle^2$, 
and $\langle \mathbf{S} \rangle=0$ when $B=0$, so that $\chi_{m}^{(i)}\propto\langle \mathbf{S}^2\rangle$.
Therefore, at finite temperature $\chi_{m}\propto \sum_{i} \langle\mathbf{S}^2\rangle \exp(-\beta\varepsilon_{i})$.
The $\langle\mathbf{S}^2\rangle$ values of the first six eigenstates are 0.251, 0.251, 0.269, 0.166, 0.166, 0.158 from low to high eigenvalues respectively. 
The third eigenstate, with the highest $\langle\mathbf{S}^2\rangle$ among the six $S=0$ states, is only $0.09\meV$ ($1.04\K$) higher than the doubly degenerate ground states.
This leads to the small peak of $\chi_{m}$ at $T\sim3\K$.
On the other hand, the three higher eigenstates with a gap about $1.18\meV$ ($13.7\K$) above the ground states have lower $\langle\mathbf{S}^2\rangle$ than the three lower eigenstates.
This leads to the small valley in $\chi_{m}$ at $T\sim25\K$.
Then, above $T\sim50\K$, a rise in $\chi_{m}$ appears as temperature increases.
This is because, as temperature increases, more $S=1$ states contribute to an increase in $\langle\mathbf{S}^2\rangle$.
Note that the antiferromagnetic character is robust for all the temperature region up to $300\K$, so that $\chi_{m}$ does not follow the paramagnetic behavior $\chi_{m}\propto 1/T$.

\begin{figure}
\includegraphics[width=1\columnwidth]{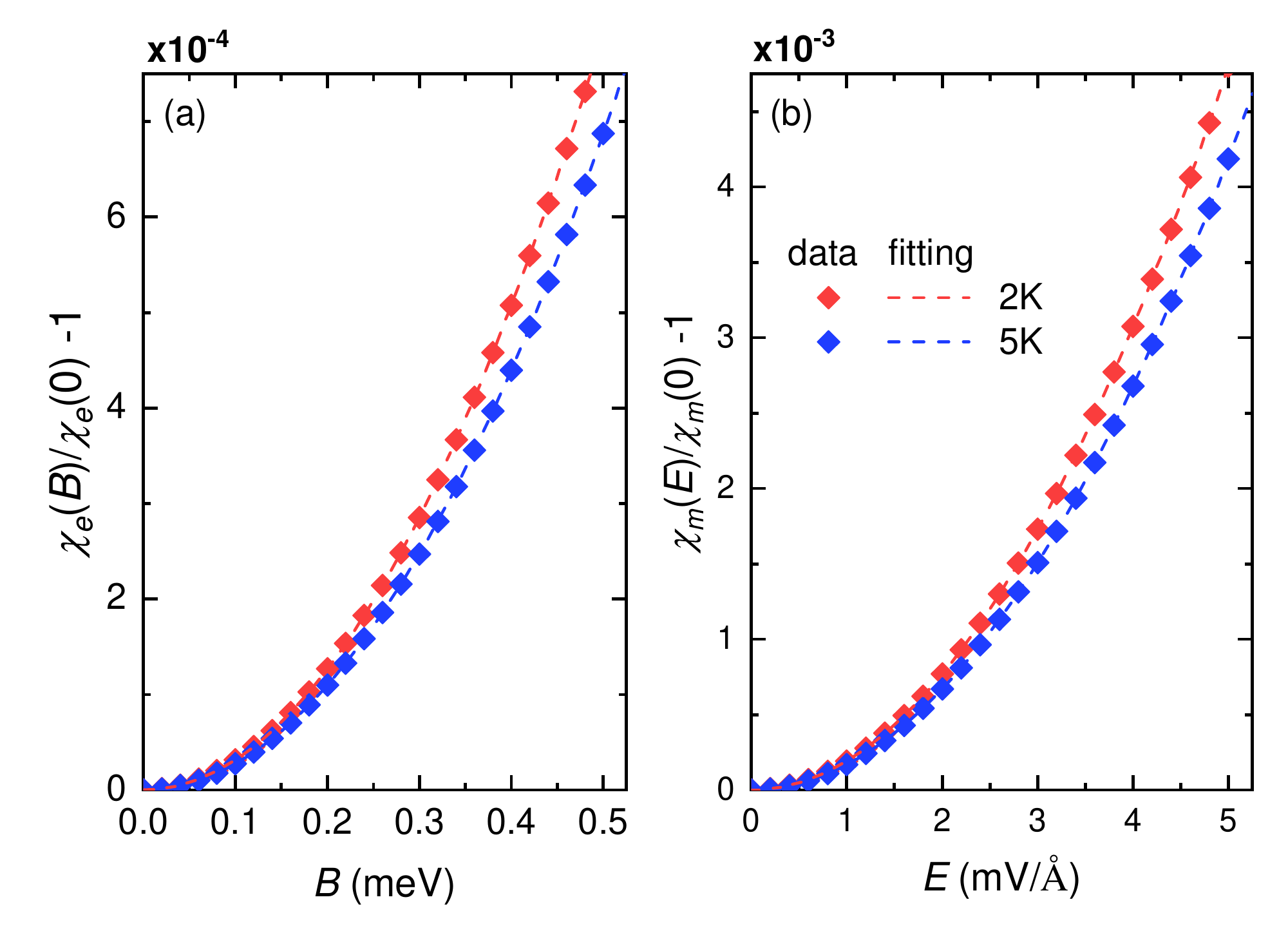} 
\caption{At $2\K$ and $5\K,$ (a) the change of electric susceptibility as a function of magnetic field $B$ and (b) the change of magnetic susceptibility as a function of electric field $E$. Solid diamonds are the data and Dashed lines are the results of quadratic fitting.}
\label{fig:thermal-x} 
\end{figure}

To further investigate the magnetoelectric coupling, we obtained the change of $\chi_{e}$ as a function of $B$ and the change of $\chi_{m}$ as a function of $E$, shown in Fig.~\ref{fig:thermal-x}. 
The fitted dotted lines show a robust quadratic relation, so that $\chi_{e}\sim B^{2}$ and $\chi_{m}\sim E^{2}$.
Since the magnetoelectric coupling originates from the dipole moment term, which involves the cross product of two spins, 
the quadratic relation is the leading order, with zero linear term according to linear response theory.
Further, the quadratic relations mean that the inversion of magnetic/electric fields leaves invariant $\chi_{e}$/$\chi_{m}$. 
It is also consistent with the DFT result that flipping spins leaves the total dipole moment invariant.

The magnitude of magnetoelectric coupling from the quantum spin model is much smaller than that found from the DFT calculations.
The reason is the quantum spin in a finite system.
In contrast to frustrated systems in solids where magnetic spins
are regarded as classical spin vectors, spins of frustrated systems
in molecular magnets often exhibit their quantum nature. 
In solids, spin vectors can rotate continuously with external fields, since the system is gapless. 
In a quantum spin system, 
there is a gap between different quantum spin states and the magnitude of the gap is positively correlated the the magnitude of the exchange interaction, DM interaction, and magnetic anisotropy. 
Once the energy of the external field is much smaller than the gap, the response is limited.

\subsection{Implication  for experimental measurements}

Fig.~\ref{fig:thermal-x}(b) shows the predicted change in spin susceptibility for experimentally accessible $E$ fields. 
The fractional change in magnetic susceptibility at $2\K$ for 
$E = 0.10 \,\textrm{V/\AA}$ (or 1 MV/m) is
\begin{equation}
\frac{\Delta \chi_{m}(E)}{\chi_m(0)} \approx 3\times 10^{-6}. 
\end{equation}
Although the change is very small, it is within the range of modern high sensitivity techniques for measuring radio frequency susceptibility \cite{Sirusi}. 
For typical  experimental applied field strengths  of order of 
 $3 \times 10^5 \, \textrm{V} \, \textrm{m}^{-1}$, the fractional change in the magnetic susceptibility is 
 ${\Delta \chi_m(E)}/{\chi_m(0)} \approx 2\times 10^{-7}$ which is comparable to experimental capabilities of the order of $ 1 \times 10^{-7}$ in the relevant temperature range.
 
 It is also significant that the dependence of the change in magnetic susceptibility on electric field strength is quadratic, as shown in Fig.\ref{fig:thermal-x}.  The absence of a linear electric effect is due to the lack of large strain dependence. 
The magnetoelectric effect is caused by a superexchange interaction via Mn-Te-Mn, or symmetric striction.  
Because of the quadratic dependence on field strength, experiments should be designed for the highest possible values of $E$ within limitations imposed by electrical breakdowns of sample cell materials and thermal bonding agents used for the samples.

\section{Conclusion}

In summary, we first investigated the magnetic properties of the crystalline phase of 
$\mathrm{Mn}_{4}\mathrm{Te}_{4}(\mathrm{P}\mathrm{Et}_{3})_{4}$ based on first principles calculations.
Each Mn has a $S=5/2$ high spin state.
The antiferromagnetic coupling leading to frustrated spins and the non-collinear DM interaction as well as the magnetic anisotropy was identified and quantized.
A non-zero electric dipole moment was obtained in non-collinear spin configurations based on Berry phase calculations.
The magnitude of the dipole moment follows the formula $\sim\hat{\mathbf{e}}_{ij}\times(\mathbf{S}_{i}\times\mathbf{S}_{j})$.
So that the electric dipole is coupled with non-collinear magnetic moment and we thus found the DM-induced ME effect in the single molecular scale.
After parameterized the spin-spin Hamiltonian, we studied the quantum spin model based on the eigenvalues and eigenstates found by the direct diagonalization of the Hamiltonian.
The magnetic susceptibility $\chi_{m}$ is changed by the electric field $E$ and the electric susceptibility  $\chi_{e}$  is changed by the magnetic field $B$, though the change is small.
Further studies showed quadratic relations between both $\chi_{m}$ and $E$, and  $\chi_{e}$ and $B$, respectively.
Such ME effect is expected to be observable in experiments.

\begin{acknowledgements}
This work is supported as part of the Center for Molecular Magnetic Quantum Materials, an Energy Frontier Research Center funded by the U.S. Department of Energy, Office of Science, Basic Energy Sciences under Award No. DE-SC0019330. 
Computations were performed at NERSC 
and the University of Florida Research Computer Center.
\end{acknowledgements}

\end{document}